\documentclass{aa} 
\usepackage{graphicx}
\usepackage{txfonts}
\usepackage{natbib}

\renewcommand{\d}{\mathrm{d}}
\newcommand{\e}{\mathrm{e}}
\newcommand{\be}{\begin{equation}}
\newcommand{\ee}{\end{equation}}
\renewcommand{\d}{\mathrm{d}}

\begin{document} 
 
\title{Searching dark-matter halos in the GaBoDS survey}

\author{Matteo Maturi\inst{1,2,3}
  \and Mischa Schirmer\inst{4}
  \and Massimo Meneghetti\inst{1,5}
  \and Matthias Bartelmann\inst{1}
  \and Lauro Moscardini\inst{3,6}
}

\institute{Zentrum f\"ur Astronomie, ITA, Universit\"at 
   Heidelberg, Albert-\"Uberle-Str.~2, 69120 Heidelberg, Germany\\ 
   \and
   Dipartimento di Astronomia, Universit\`a di Padova, Vicolo 
   dell'Osservatorio 2, 35120 Padova, Italy \\
   \and
   Dipartimento di Astronomia, Universit\`a di Bologna, Via 
   Ranzani 1, 40127 Bologna, Italy\\
   \and
   Isaac Newton Group of Telescopes, Calle Alvarez Abreu 70, 
   38700 Santa Cruz de La Palma, Spain\\
   \and
   INAF - Osservatorio Astronomico di Bologna, Via Ranzani 1, 40127 
   Bologna, Italy\\
   \and
   INFN, Sezione di Bologna, viale Berti Pichat 6/2,
   I-40127 Bologna, Italy}

\date{\emph{Astronomy \& Astrophysics, submitted}} 

\abstract
{
 We apply the linear filter for the weak-lensing signal of dark-matter
 halos developed in Maturi et al. (2005) to the cosmic-shear data
 extracted from the Garching-Bonn-Deep-Survey (GaBoDS).}
{
 We wish to search for dark-matter halos through weak-lensing
 signatures which are significantly above the random and systematic
 noise level caused by intervening large-scale structures.}
{
 We employ a linear matched filter which maximises the signal-to-noise
 ratio by minimising the number of spurious detections caused by the
 superposition of large-scale structures (LSS). This is achieved by
 suppressing those spatial frequencies dominated by the LSS
 contamination.}
{
 We confirm the improved stability and reliability of the detections
 achieved with our new filter compared to the commonly-used aperture
 mass (Schneider, 1996; Schneider et al., 1998) and to the aperture
 mass based on the shear profile expected for NFW haloes (see
 e.g.~Schirmer et al., 2004; Hennawi \& Spergel, 2005). Schirmer et
 al.~(2006) achieved results comparable to our filter, but probably
 only because of the low average redshift of the background sources in
 GaBoDS, which keeps the LSS contamination low. For deeper data, the
 difference will be more important, as shown by Maturi et al. (2005).}
{
 We detect fourteen halos on about eighteen square degrees selected
 from the survey. Five are known clusters, two are associated with
 over-densities of galaxies visible in the GaBoDS image, and seven
 have no known optical or X-ray counterparts.
}

\maketitle

\section{Introduction}

Claims have been made in the literature that dark mass concentrations were
significantly detected through their weak-lensing signal \citep[see
e.g.][]{ER00.1,UM00.1,MI02.1,ER03.1}. If confirmed, these detections were
extremely exciting because they showed that large, dark mass concentrations
could exist which for some reason failed to emit light, either as stellar
light from galaxies or X-ray emission from hot intracluster plasma. Since the
baryon fraction in clusters should faithfully reflect the cosmic mixture of
baryonic and dark mass \citep{ET03.1,ET06.1}, detections of truly dark
cluster-sized halos could shed doubt on our understanding of the formation of
non-linear cosmic structures.

Obviously, gravitational lensing is the only method able to detect
dark mass concentrations. This strength is also its weakness: all
density inhomogeneities projected along the line-of-sight cause
gravitational light deflection, thus any detection of a lensing signal
possibly due to a dark-matter halo is affected to some degree by the
large-scale structures projected into the observed
field. Superpositions of large-scale structures can create signals
mimicking dark-matter halos.

Commonly used methods for detecting halos through weak lensing, in particular
the aperture mass \citep{SC96.2,SC98.2}, are optimised for measuring the total
projected mass enclosed within an aperture, and for suppressing the
cross-correlation of the signal in neighbouring apertures.

A clean separation between the signals of halos and large-scale
structures is strictly impossible because the large-scale structure
can be considered as composed of halos with a broad mass
spectrum. Unambiguous noise suppression can thus not be
achieved. However, we have argued in \citep{MAT04.2} that the
weak-lensing power spectrum obtained from the \emph{linearly} evolved
dark-matter power spectrum can reasonably be considered as a noise
contribution against which the weak-lensing signal of halos can be
filtered. The underlying assumption is that those halos which we are
searching for do in fact create the non-linear power spectrum.

Using simulations, we demonstrated that the linear filter which follows
uniquely from this noise model and the assumption that dark-matter halos on
average have an NFW density profile does indeed perform as expected
\citep{MAT04.2}. It suppresses the noise from large-scale structures
substantially, thus considerably reducing the number of spurious detections,
and is much less sensitive than the aperture mass against changes in the
filter scale. Removing the LSS contamination is less important for relatively
shallow observations, i.e.~for average source redshifts
$z_\mathrm{s}\lesssim1$, but it is fundamental for deep observations in which
the integrated contribution of the matter along the line-of-sight becomes
non-negligible.

We now apply this filter to the weak-lensing data obtained from the
Garching-Bonn-Deep-Survey \citep[hereafter \textit{GaBoDS},
see][]{SC03.2}. The purpose of this study is two-fold. First, we wish to
compare the halo detections with the new filter and with the aperture mass,
and second, we want to assess the reliability of the halo detections and their
stability against changes in the filter scale.

We emphasise that we do not want to devaluate the aperture mass. It
was shown to have many desirable properties for the measurement of
cosmic shear. However, its construction for a different purpose than
halo detection motivates us to search for an alternative
measure. Ultimately, our filter can be seen as a variant of the
aperture mass based on a considerably narrower angular weight
function, which is characterised by the assumed halo density profile
and the noise model.

Throughout, we assume that dark matter halos have density profiles of NFW
shape and use the angular scale radius of the profile as a free
parameter. Moreover, we compute the weak-lensing power spectrum of
\emph{linear} large-scale structures adopting a $\Lambda$CDM cosmological
model, with $\Omega_0=0.3$, $\Omega_\Lambda=0.7$, $\sigma_8=0.9$ and $h=0.7$.

The paper is structured as follows: Sect.~2 briefly summarises the
survey characteristics; Sects.~3 and 4 describe the aperture mass and
our optimal filter, respectively; and Sect.~5 presents the analysis of
the data sample. Our conclusions are presented in Sect.~6.

\section{The GaBoDS survey\label{sec:survey}}

The Garching-Bonn Deep Survey collects observations obtained with the Wide
Field Imager (WFI) mounted on the MPG/ESO 2.2m telescope. The survey covers
$\sim19$ square degrees in the $R$ band and consists of shallow
($4-7\,\mbox{ksec}$, $9.6\,\mbox{deg}^2$), medium ($8-11\,\mbox{ksec}$,
$7.4\,\mbox{deg}^2$) and deep ($13-56\,\mbox{ksec}$, $2.6\,\mbox{deg}^2$)
exposures taken under sub-arcsecond seeing conditions. To a large extent, the
data constitute a virtual survey in the sense that a very substantial fraction
of it was identified in the ESO archive.

The fields are randomly distributed in the southern sky and at high Galactic
latitude, avoiding bright stars, extended foreground objects and to some
degree also known massive galaxy clusters (for more details see Schirmer et
al.~2003).

The number density of galaxies used for our weak lensing study
reflects the inhomogeneous depth of the survey and ranges from
$n=6-24\,\mbox{arcmin}^{-2}$, with an average of $\bar{n}\approx
12\,\mbox{arcmin}^{-2}$.

A comprehensive characterisation of the survey can be found in
\citep{SC03.2}, including references to its suitability for weak
lensing studies. The data reduction was performed with the
\textit{THELI} pipeline, which \citet{ER05.1} describe in detail.

\section{Aperture mass\label{sec:ap_filter}}

The aperture mass is widely used for weak-lensing studies. It is
defined as a weighted integral of the local convergence,
\begin{equation}\label{eq:map}
  M_\mathrm{ap}(\vec\theta)=\int\d^2\theta'\,\kappa(\vec\theta')
  U(\vec\theta-\vec\theta')\;,
\end{equation}
where the aperture is chosen to be circular and $U$ is a \emph{compensated}
axially symmetric weight function, i.e.~it satisfies
\begin{eqnarray}
  U(\theta)& = & U(|\vec\theta|) \\
           & = &\int_0^\theta\d\theta'\,\theta'\,U(\theta')=0\;.
\end{eqnarray}

\begin{figure}[ht] 
  \includegraphics[width=\hsize]{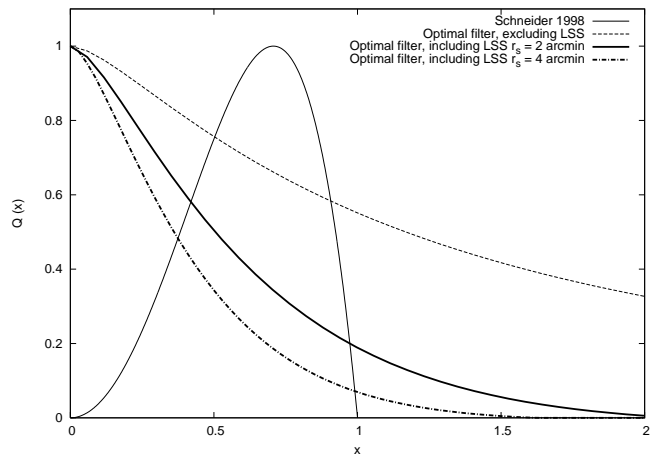} 
\caption{Comparison between the filter profiles. Different line types refer
  to the conventional aperture mass (solid line), the optimal filter described
  in this paper but without the large-scale structure contribution to the
  noise power spectrum (dashed line), the full optimal filter for $r_s=2'$ and
  $r_s=4'$ (heavy solid and heavy dash-dotted lines, respectively). The
  large-scale structure power spectrum modifies the filter shape according to
  the filter scale by lowering its sensitivity on large scales where weak
  lensing by large-scale structure dominates.}
\label{fig:compare_filt} 
\end{figure}  

Since the convergence is not directly measurable, Eq.~(\ref{eq:map})
is conveniently expressed in terms of the shear component
$\gamma_\mathrm{t}$ tangential with respect to the aperture centre
$\vec\theta$ on the sky, and can be written in the form
\begin{equation}\label{eq:mapt}
  M_\mathrm{ap}(\vec\theta)=
  \int\d^2\theta'\,\gamma_\mathrm{t}(\vec\theta')
  Q(|\vec\theta-\vec\theta'|)\;,
\end{equation}
where the function $Q$ is related to $U(\theta)$ through
\begin{equation}
  Q(\theta)\equiv\frac{2}{\theta^2}\int_0^\theta\d\theta'\,
  \theta'\,U(\theta')-U(\theta)\;.
\end{equation}

The function $Q$ is usually chosen to have a compact support, which is
adequate for the finite data fields on which halos are sought. It is
also typically chosen to suppress the halo centre because there the
tangential component of the ellipticity is not defined, the
weak-lensing approximation may break down, and cluster galaxies may
prevent the shapes of background galaxies from being precisely
measured.

\cite{SC98.2} proposed a polynomial shape for $Q$,
\begin{equation}\label{eq:ap_pol}
  Q_\mathrm{pol}(x)=\frac{(1+l)(2+l)}{\pi\theta^2}\,
  x^2(1-x^2)^l\mathrm{H}(1-x)\,,
\end{equation}
where $\mathrm{H}$ is the Heaviside step function, $l$ is a free parameter
usually set to $l=1$, $x\equiv\theta/\Theta$ and $\Theta$ is the radius where
$Q$ vanishes.

These filter profiles do not minimise the variance of the estimate
$M_\mathrm{ap}$, and their aperture can be arbitrarily chosen by the
observer, leading to unstable results which sensitively depend on the
filter scale \citep{MAT04.2}. It was designed for cosmic shear
statistics, but it is widely used in the literature for halo searches.

In recent years, other function types have been introduced for $Q$ in order to
maximise the signal-to-noise ratio.
\cite{SC98.2} showed that this is achieved if $Q$ mimics the tangential shear
profile of the lens. For example, \cite{SC04.2,HE04.1} and \cite{PA03.2} use
functions which approximate the expected shear profile of a NFW halo.

Specifically, \cite{SC04.2} used the following parameterisation
\begin{equation}\label{eq:ap_tanh}
  Q_\mathrm{tanh}(x)=\frac{1}{1+\e^{6-150x}+\e^{-47+50x}}
  \frac{\tanh(x/x_\mathrm{c})}{x/x_\mathrm{c}}\,,
\end{equation}
where $x_\mathrm{c}$ defines how peaked the profile is in analogy to
the scale radius of the NFW profile. The exponential pre-factor damps
the filter within its inner and outermost 10\%.

However, all these functions have some disadvantage since they neglect
the presence of the cosmic shear from large-scale structures which
dominates the halo lensing signal above some scale and thus acts as an
intense source of noise.

\section{Optimal filter\label{sec:opt_filter}}

\begin{table}
  \caption{GaBoDS fields rejected because of an excessive $B$-mode
    contamination. Each column lists the names of the rejected fields, where
    the selection was performed for each filter and all scales because
    they determine the sensitivity for the residual $B$-mode.
  \label{tab:rejected_F}}
\begin{center}
  \begin{tabular}{|c|cc|cc|}
    \hline
    $M_{ap}$ 4' &$M_{oap}$ 10' &$M_{oap}$ 20' &$A_{opt}$ 2' &$A_{opt}$ 4'\\
    \hline
    A1347\_P3 &-      &-         &-      &-\\
    -         &-      &A1347\_P4 &-      &-\\
    B8m3      &-      &-         &-      &-\\
    -         &-      &B8p0      &-      &B8p0\\
    B8p2      &-      &-         &-      &-\\
    -         &-      &B8p3      &-      &B8p3\\
    -         &-      &C0400     &-      &C0400\\
    C04p3     &C04p3  &C04p3     &C04p3  &C04p3\\
    CL1119    &-      &-         &-      &-\\
    -         &-      &Cl1301    &-      &-\\
    -         &-      &-         &-      &DEEP3b\\
    FDF       &-      &-         &-      &-\\
    -         &Pal3   &Pal3      &Pal3   &Pal3\\
    SGP       &-      &-         &-      &-\\
    \hline
  \end{tabular}
\end{center}
\end{table}

\begin{figure*}[t]
  \includegraphics[width=0.32\hsize]{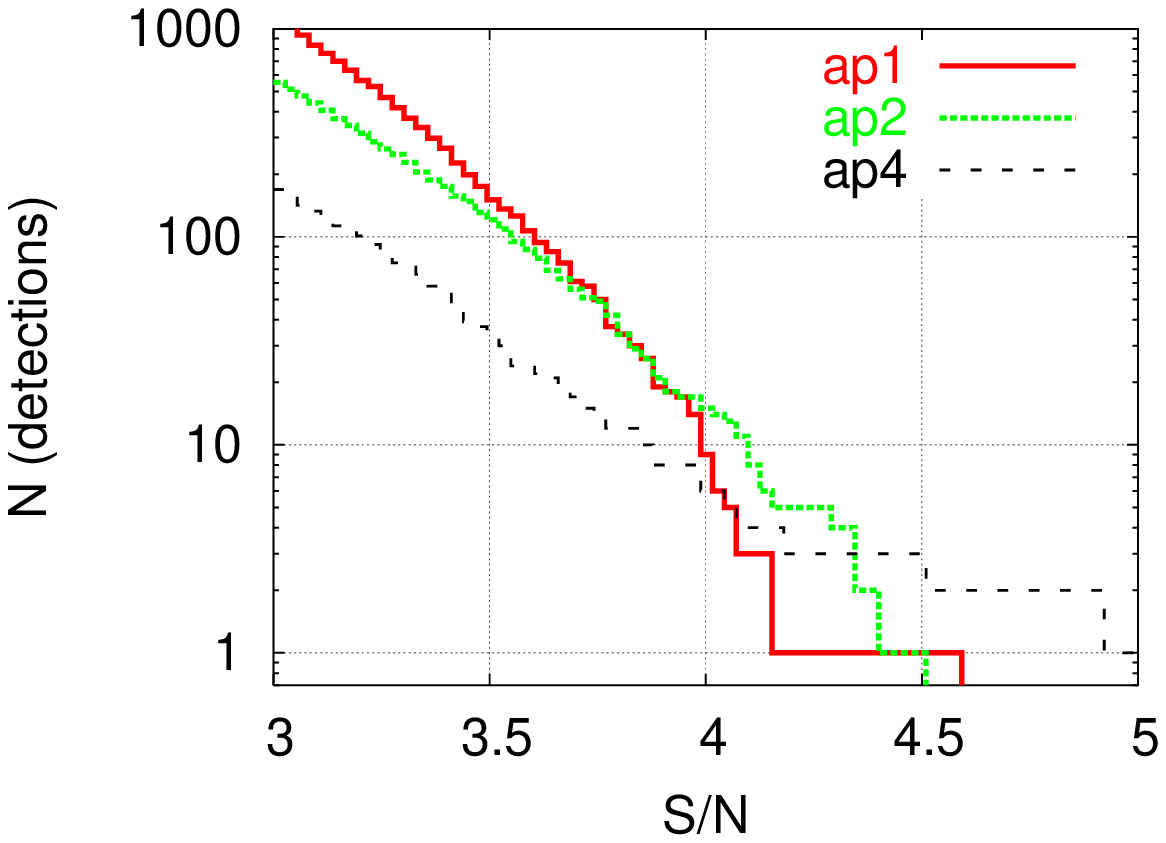}\hfill
  \includegraphics[width=0.32\hsize]{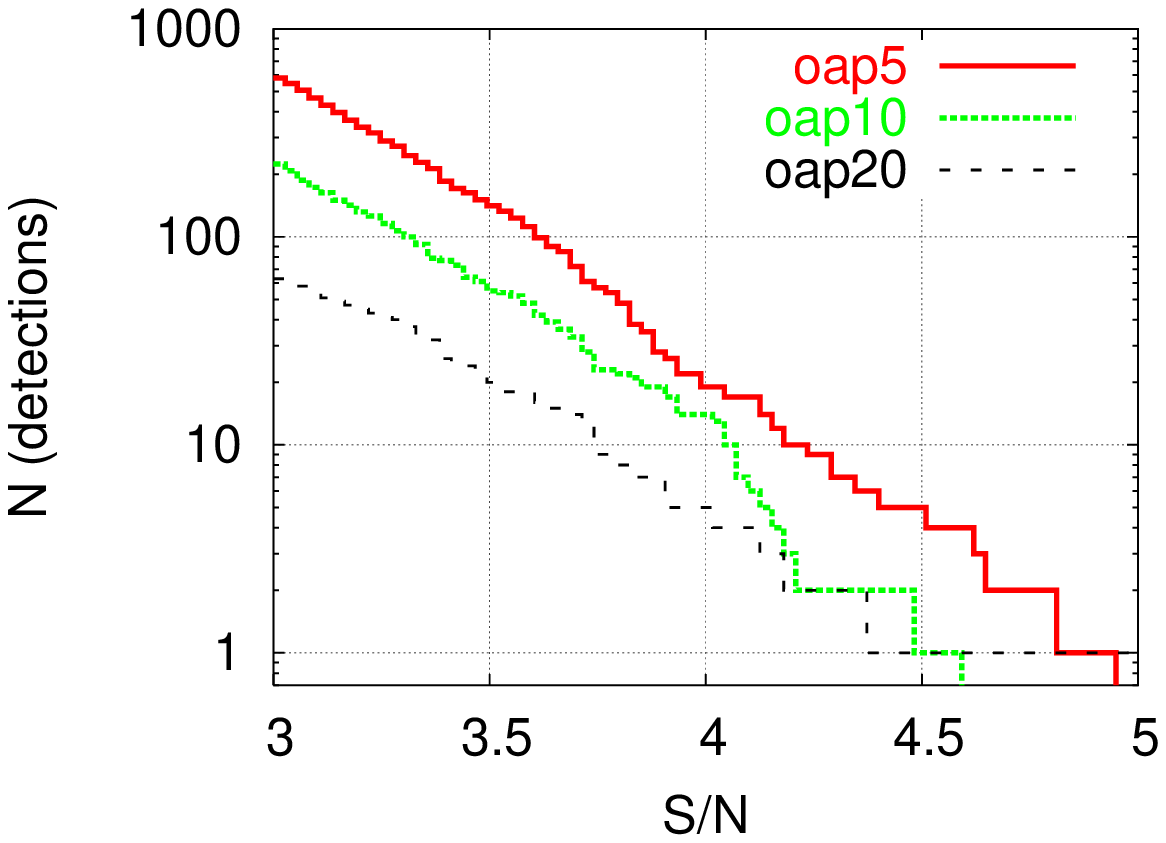}\hfill
  \includegraphics[width=0.32\hsize]{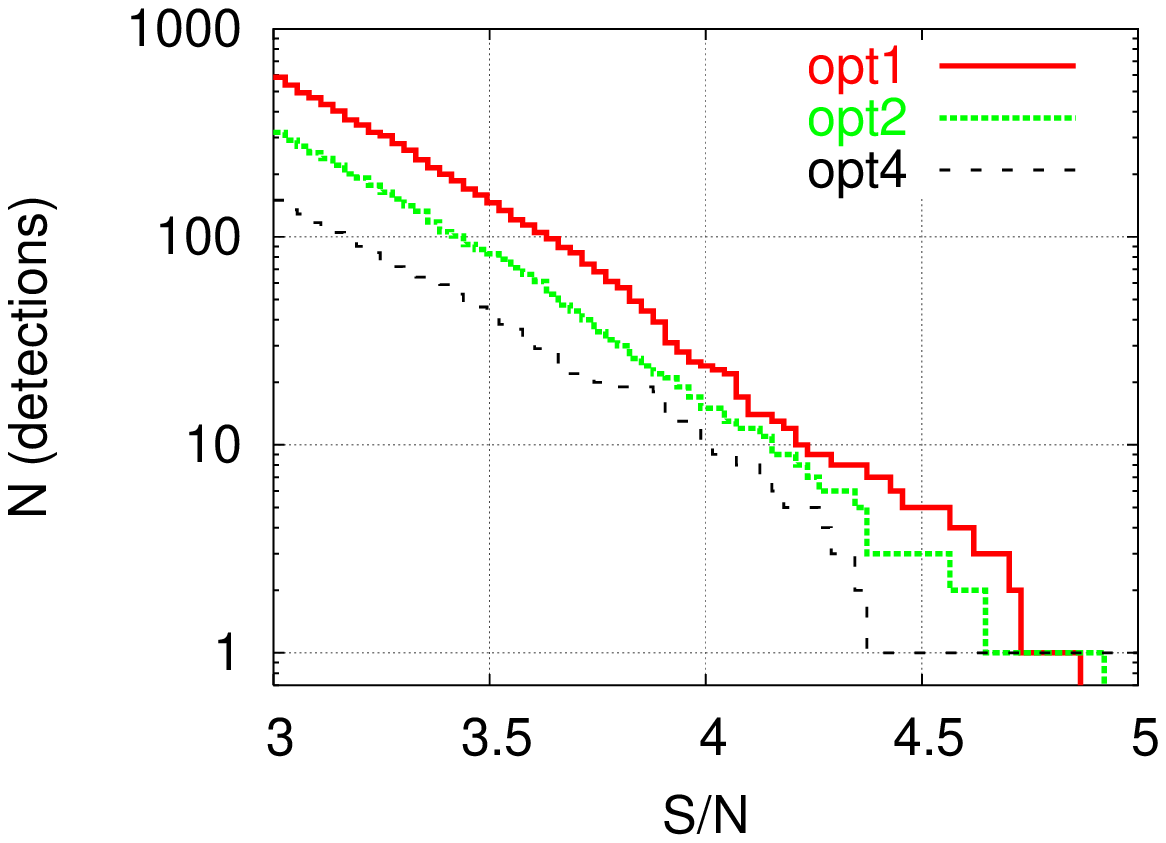}
  \includegraphics[width=0.32\hsize]{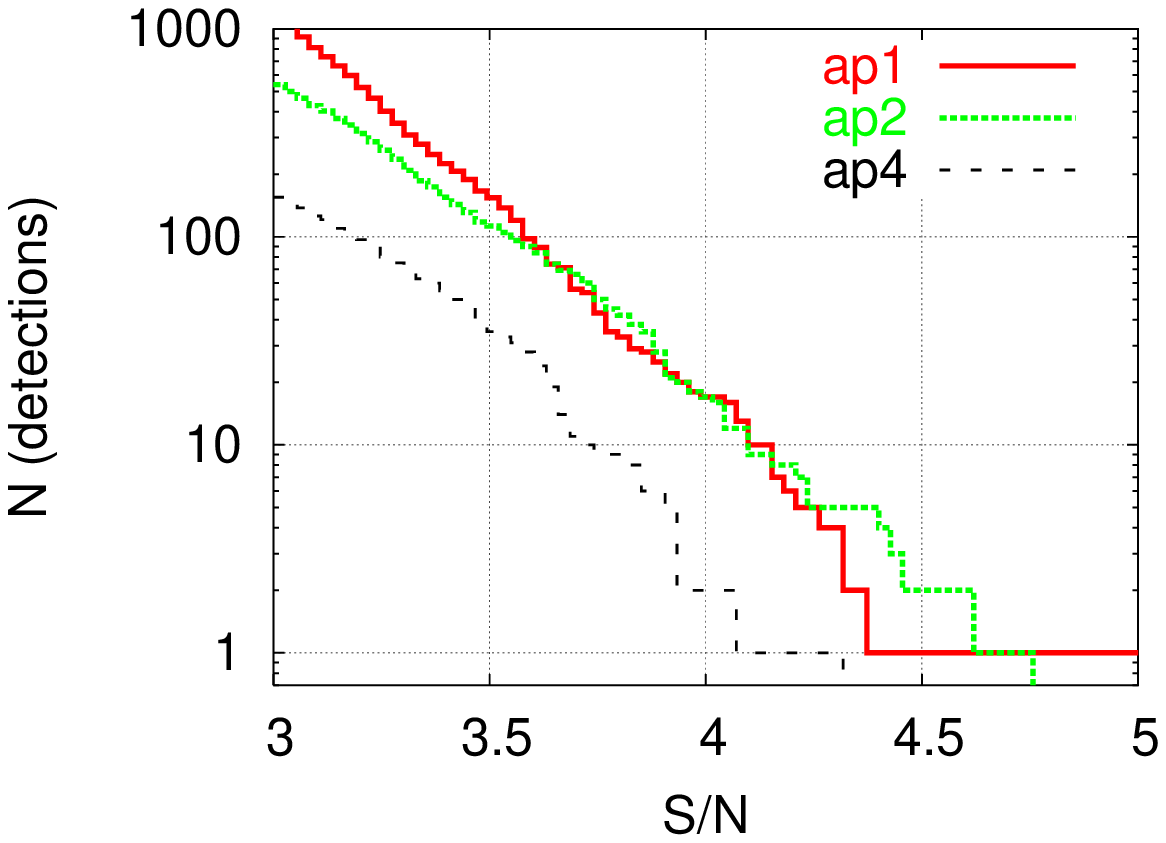}\hfill
  \includegraphics[width=0.32\hsize]{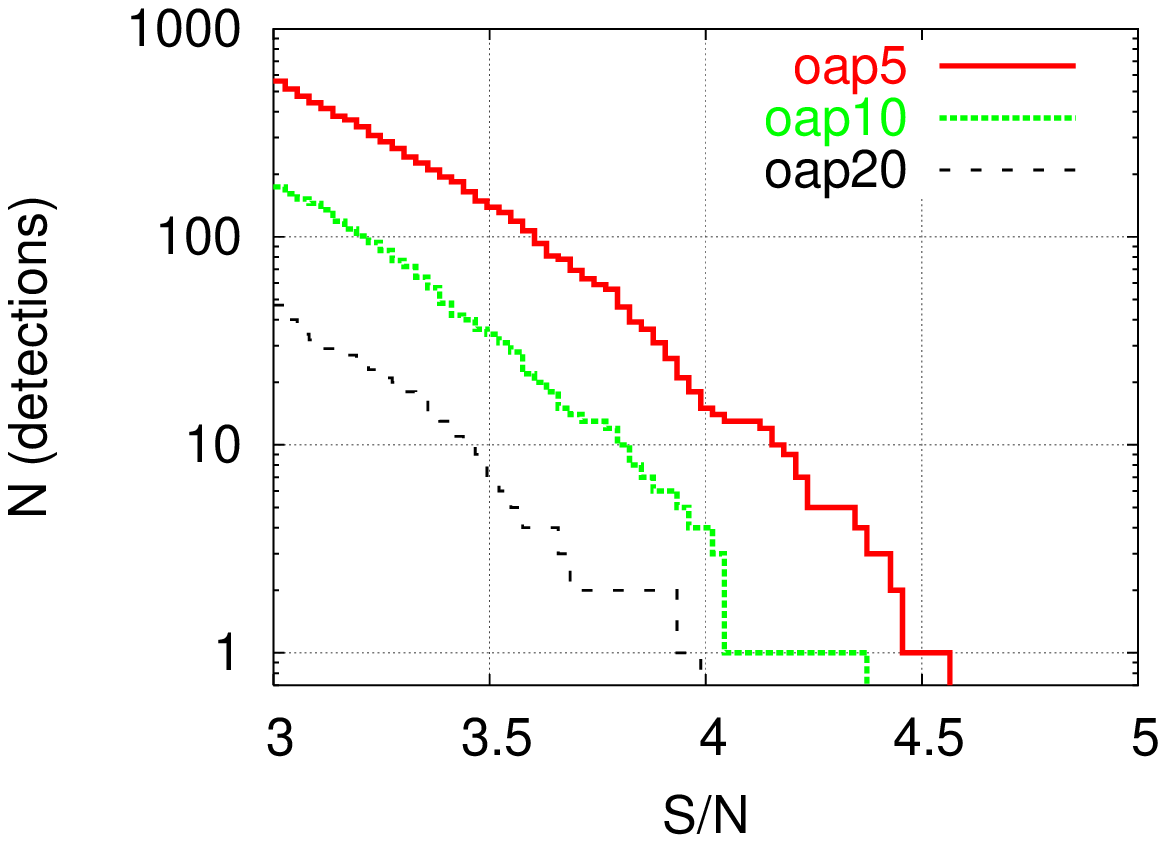}\hfill
  \includegraphics[width=0.32\hsize]{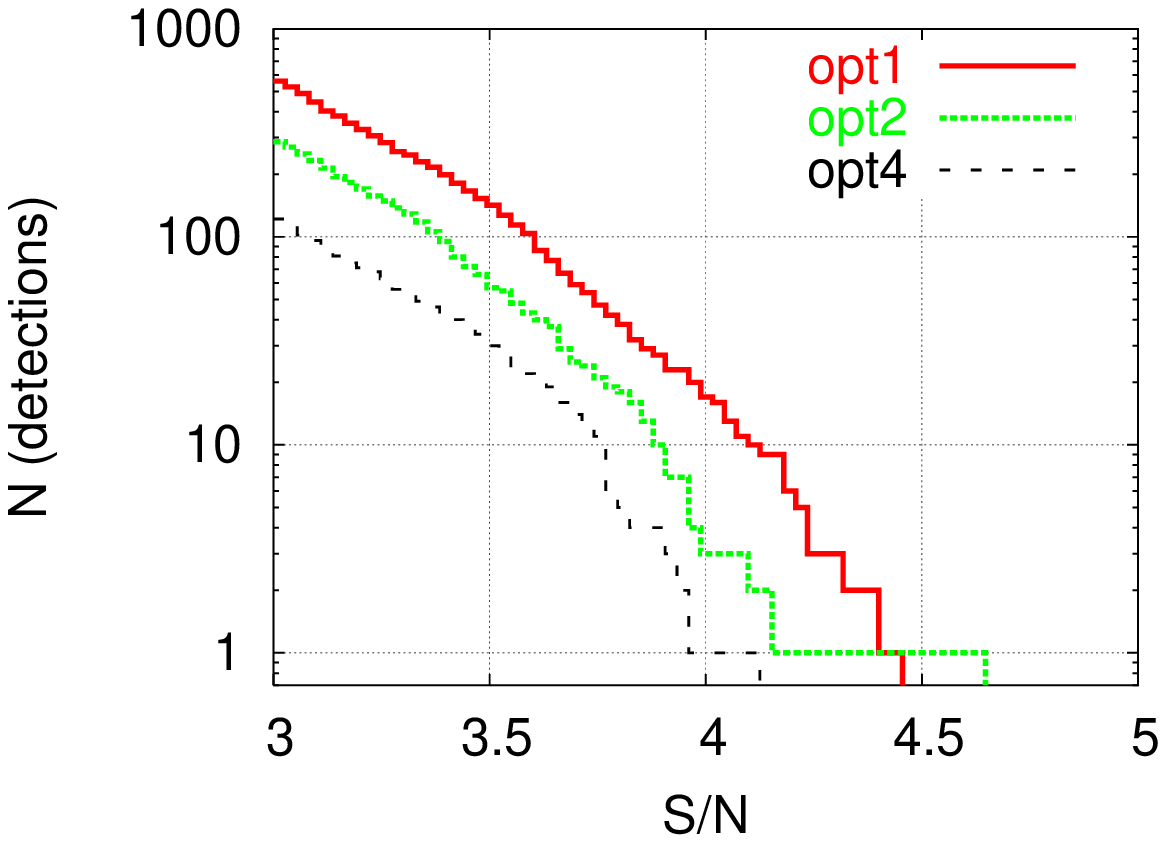}
\caption{Cumulative distribution of the S-detections (upper panels)
   and of the N-detections (bottom panels) as a function of their
   signal-to-noise ratio. Three filter functions were used: the
   aperture mass $M_\mathrm{ap}$ as described by \cite{SC96.2} (left
   panels), the aperture mass $M_\mathrm{oap}$ as defined by
   \cite{SC04.2} (middle panels), and the optimal filter
   $A_\mathrm{opt}$ developed by \cite{MAT04.2} (right panels). Three
   different scales were used: $1'$, $2'$ and $4'$ for
   $M_\mathrm{ap}$ and $A_\mathrm{opt}$, and $5'$, $10'$ and $20'$ for
   $M_\mathrm{oap}$. The number of spurious N-detections is high with
   all filters. However, only with $M_\mathrm{oap}$ and
   $A_\mathrm{opt}$ have the distributions of S-detections a tail
   extending towards high signal-to-noise ratio for all filter scales,
   exceeding the number of noise detections. A comparison between the
   distributions of the S-detections and of the N-detections is more
   clearly shown in Fig.~(\ref{fig:cum_SN-N}).\label{fig:cum_SN}}
\end{figure*}

\begin{figure*}[t]
  \includegraphics[width=0.32\hsize]{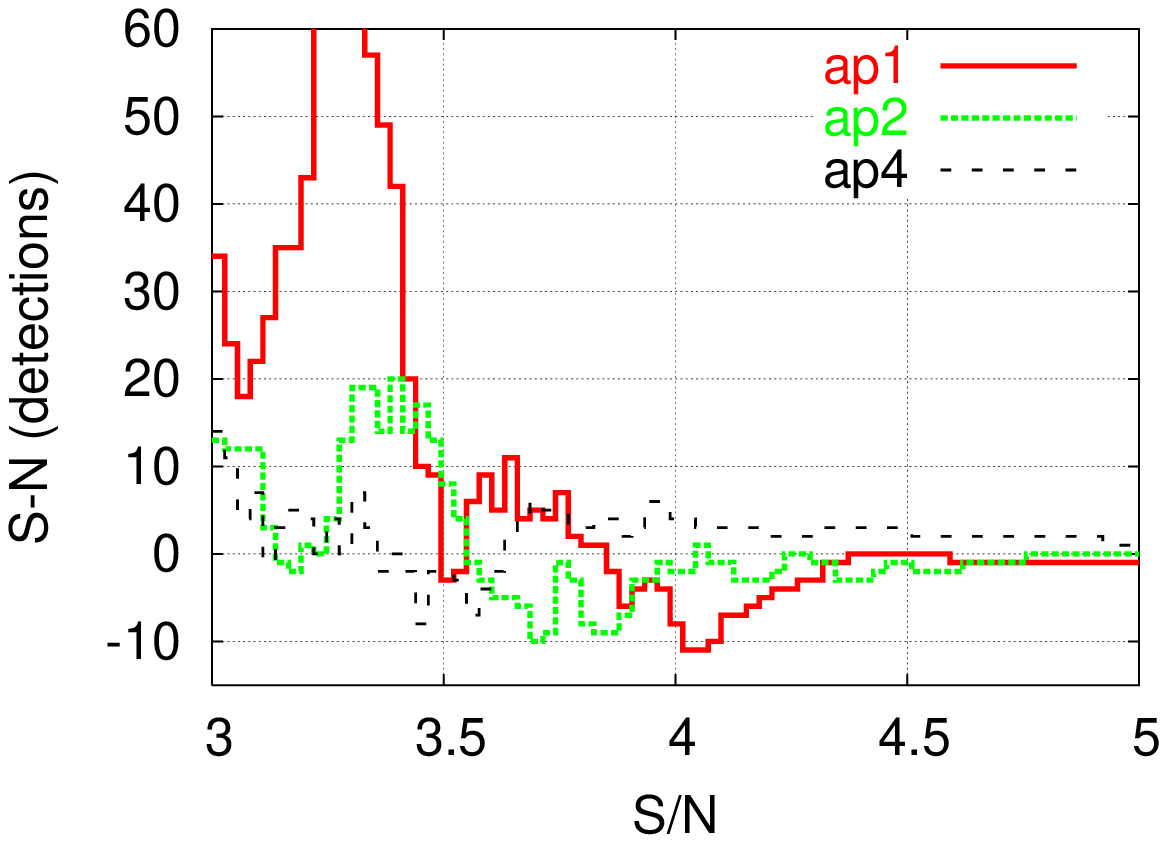}\hfill
  \includegraphics[width=0.32\hsize]{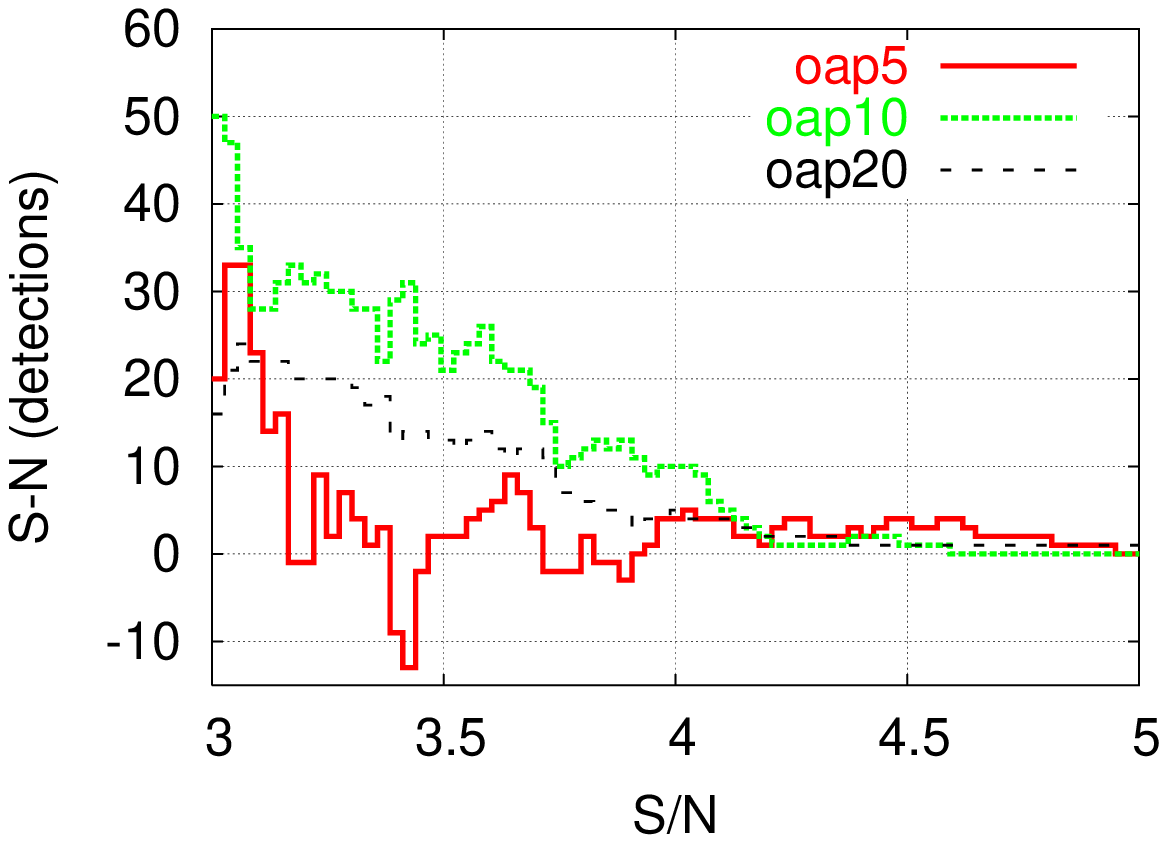}\hfill
  \includegraphics[width=0.32\hsize]{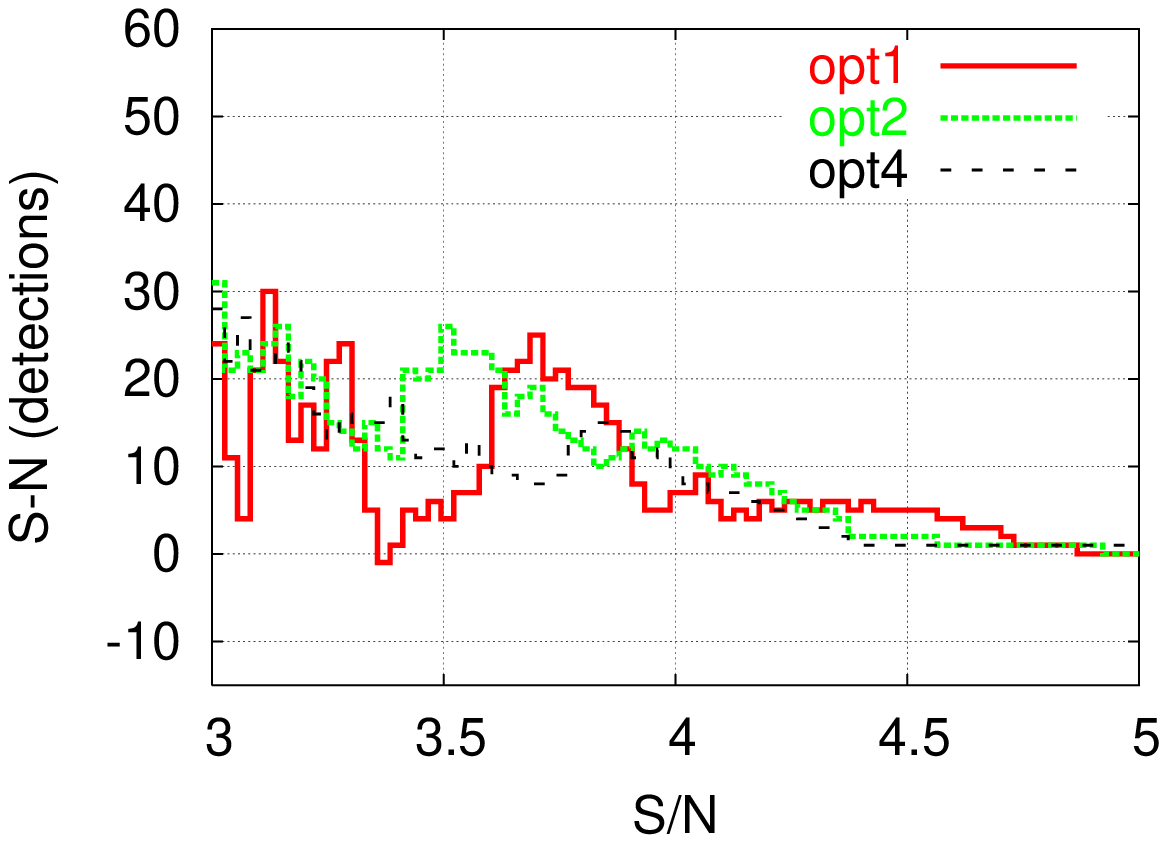}
  \includegraphics[width=0.32\hsize]{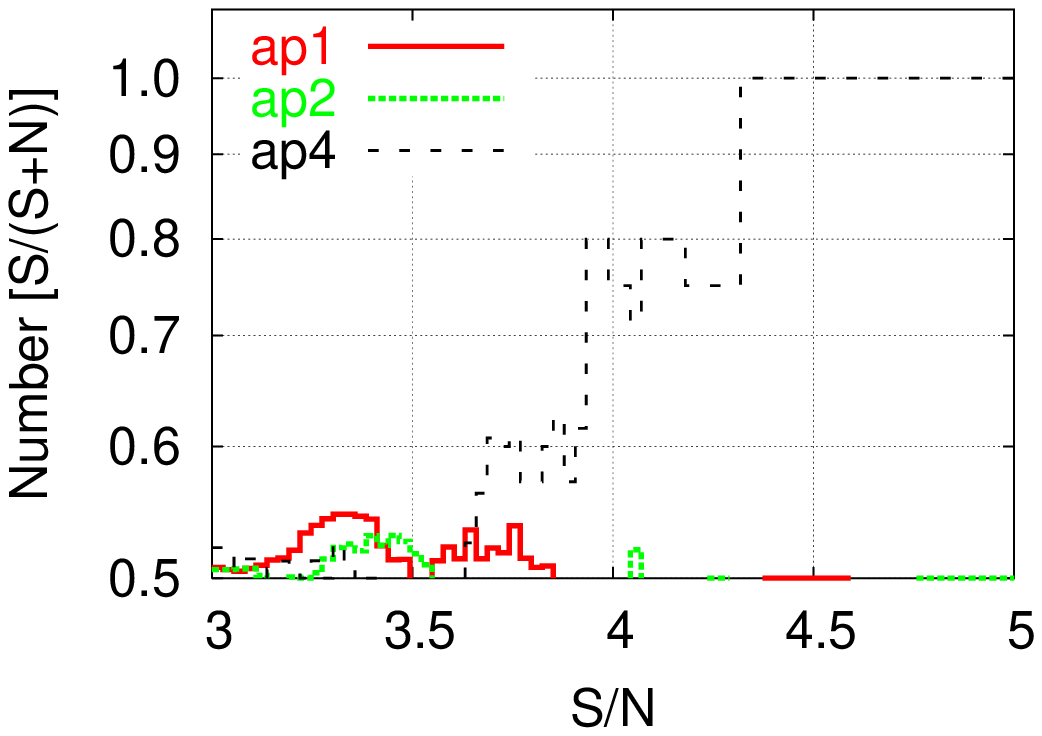}\hfill
  \includegraphics[width=0.32\hsize]{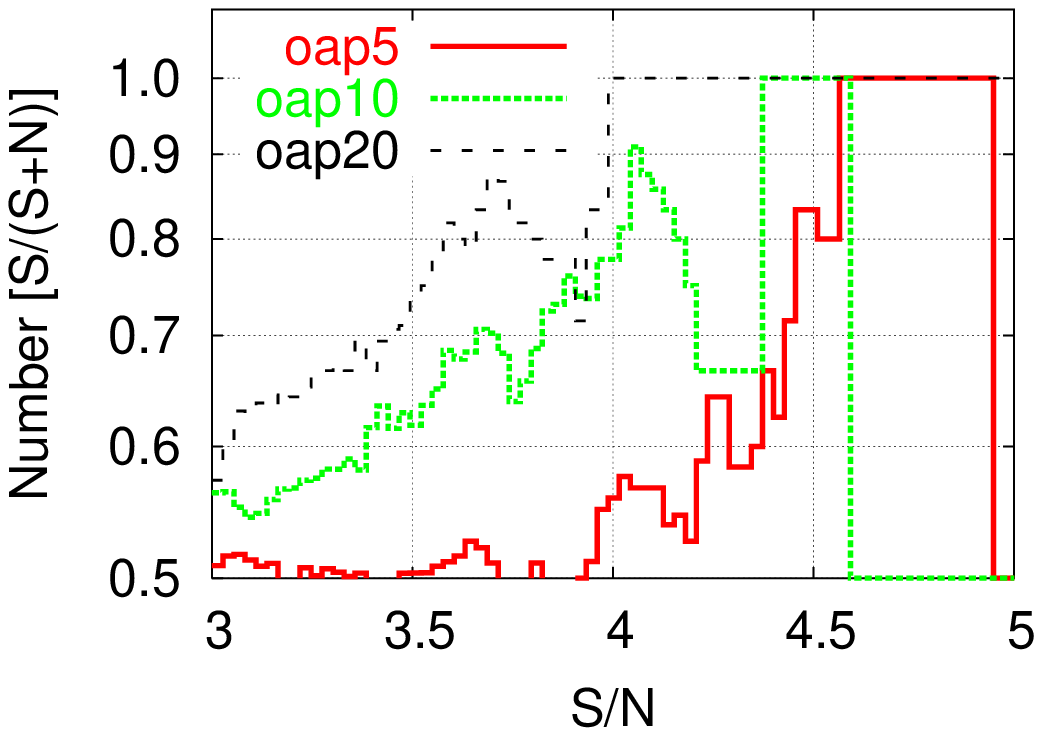}\hfill
  \includegraphics[width=0.32\hsize]{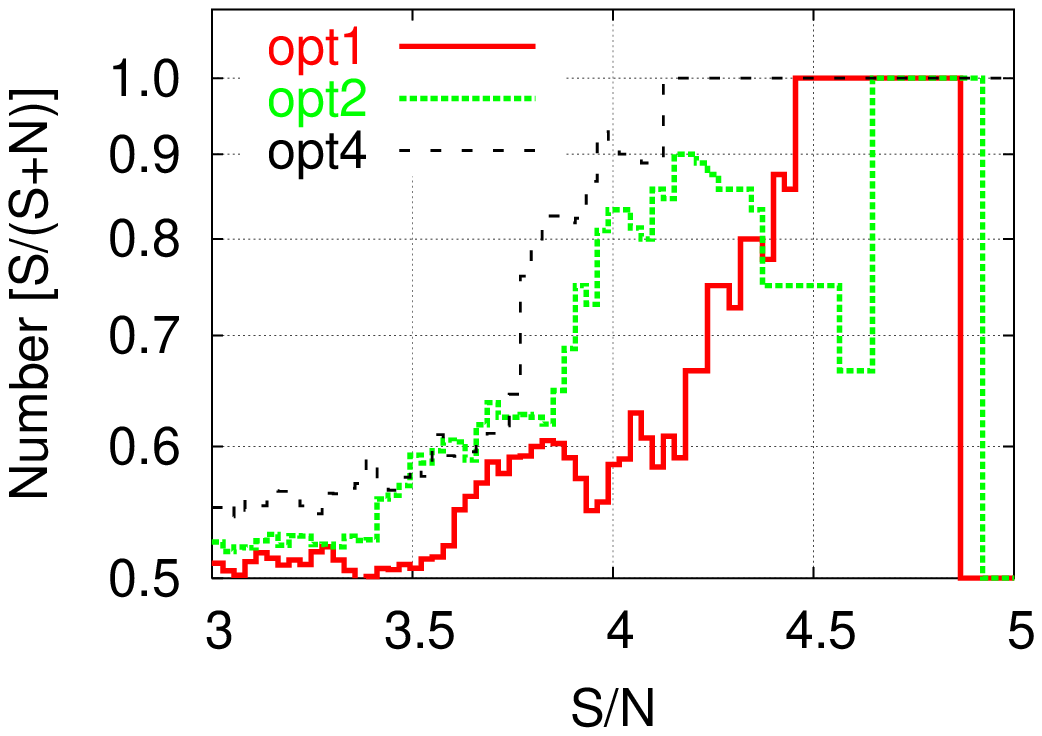}
\caption{In order to more clearly compare the S- and N-distributions
  for each filter, we plot their difference (upper panels) and the
  ratio between the S-distribution and the sum of the S- and
  N-distributions (bottom panels). The original aperture mass
  $M_\mathrm{ap}$ (left panels) shows a significant contribution of
  S-detections only for an aperture of $4'$. In comparison,
  $M_\mathrm{oap}$ (middle panels) and $A_\mathrm{opt}$ (right panels)
  give a net positive contribution for large signal-to-noise ratios on
  all scales. Consequently, it is possible to define a threshold S/N
  ratio above which the detections can be considered reliable. This
  plot underlines the superior statistical properties of the two
  optimised filters compared to the aperture mass $M_\mathrm{ap}$. It
  also shows that $A_\mathrm{opt}$ yields more reliable and stable
  results than the other filters. In fact, at high signal-to-noise
  ratios, it produces the largest difference between S- and
  N-detections.}
\label{fig:cum_SN-N}
\end{figure*}

As discussed by \cite{MAT04.2}, we shall model this weak-lensing
contribution as a noise component to be filtered out. The measured
data $D(\vec\theta)$ is thus composed by the signal $S(\vec\theta)$
and some noise $N(\vec\theta)$,
\begin{equation}\label{eqn:signal} 
  D(\vec\theta)=S(\vec\theta)+N(\vec\theta)=
  A\tau(\vec\theta)+N(\vec\theta)\;,
\end{equation} 
expressing that the signal is modelled as the weak-lensing signal of a
dark-matter halo with amplitude $A$ and angular shape
$\tau(\vec\theta)$. We assume $\tau$ is axially symmetric on average,
thus $\tau(\vec\theta)=\tau(|\vec\theta|)$. Since the signal contained
in our data $D$ is the reduced shear $g$, we now identify $\tau$ with
the reduced shear expected from a given halo model, and approximate
the reduced shear by the shear $\gamma$. This is justified for
weak-lensing, $\kappa\ll1$.

The noise component $N$ is assumed to be random with zero mean and
isotropic, since to sufficient approximation the background galaxies
are randomly placed and oriented, and weak lensing by large-scale
structures is well described by an isotropic Gaussian random
field. The variances of the noise components are conveniently
described in the Fourier domain, where their correlation functions are
given by the power spectrum
\begin{equation} 
  \langle\hat N(\vec k')^*\hat N(\vec k)\rangle= 
  (2\pi)^2\,\delta(\vec k'-\vec k)P_N(k)\;.
\end{equation}  
The hats above symbols denote their Fourier transforms, and $P_N(k)$
is the noise power spectrum. Thus the noise given by the intrinsic
ellipticity of the sources combined with their finite number is
modelled by its power spectrum $P_\epsilon$, and the noise caused by
weak lensing of intervening large-scale structures is modelled by the
cosmic-shear power spectrum $P_\gamma$ derived from the linear
dark-matter power spectrum. The complete noise power spectrum is thus
\begin{equation}\label{eq:noise_spectrum}
  P_N(k)=P_\epsilon(k)+P_\gamma(k)=
  \frac{1}{2}\frac{\sigma_{\epsilon_\mathrm{s}}^2}{n_\mathrm{g}}+
  \frac{1}{2}\,P_\gamma(k)\;,
\end{equation} 
where the factor $1/2$ arises because only one (the tangential)
component of the ellipticity contributes to the measurement. The
linear filter $\Psi(\vec\theta)$ is constructed to yield an estimate
$A_\mathrm{est}$ for the amplitude $A$ of the signal at position
$\vec\theta$,
\begin{equation}\label{eq:estimate}
  A_\mathrm{est}(\vec\theta)=\int D(\vec\theta') 
  \Psi(\vec\theta-\vec\theta')\,\d^2\theta'\;,
\end{equation}   
which is unbiased, i.e.~the mean error has to vanish $b\equiv\langle
A_{\rm est}- A\rangle=0$, and minimises the measurement \emph{noise}
$\sigma^2\equiv\left\langle(A_\mathrm{est}-A)^2\right\rangle$. A
filter $\Psi$ satisfying these two conditions is found by introducing
a Lagrangian multiplier $\lambda$ and searching for the function
$\Psi$ which minimises the functional $L[\Psi]=\sigma^2+\lambda
b$. This yields
\begin{equation}\label{eq:opt_filter}
  \hat\Psi(\vec k)=\frac{1}{(2\pi)^2}\left[ 
    \int\frac{|\hat\tau(\vec k)|^2}{P_N(k)}\d^2k 
  \right]^{-1}\,\frac{\hat\tau(\vec k)}{P_N(k)}\;.
\label{eq:optfilter} 
\end{equation}
By construction, the filter $\Psi$ is thus most sensitive for those
spatial frequencies where the signal $\hat\tau$ is large and the noise
$P_N(k)$ is low \citep[for details see][]{MAT04.2}.

Throughout the paper, we assume that the signal is faithfully modelled
by the shear of an NFW profile \citep[see
e.g.][]{BA96.1,WR00.1,LI02.1,ME03.1}. Consequently, the filter is
optimised to detect the same halo shape as the optimised aperture mass
given by Eq.~(\ref{eq:ap_tanh}). We emphasise that, in contrast to the
aperture mass, this filter is not constructed from an arbitrary
compensated filter function for the convergence, but defined such that
it maximises the signal-to-noise ratio and minimises the contamination
induced by the LSS or, in principle, by any Gaussian noise with known
power spectrum. It has the further advantage of being more stable
against changes of the filter size, due to the shape control imposed
by the noise power spectrum. Thus, like the aperture mass, this
approach is based on an ordinary convolution (\ref{eq:estimate}) of
the data with a filter $\Psi$ or $Q$, but its signal-to-noise
properties differ from that of the aperture mass because they are
determined by the specific choice of the filter. The different filter
profiles are compared in Fig.~(\ref{fig:compare_filt}).

To fully exploit these properties, deep observations with a high
galaxy number density are needed, because if the shot noise is
dominant with respect to the LSS noise, its performance is expected to
be comparable to the optimised aperture mass.

\section{Analysis of the GaBoDS data}

The GaBoDS survey data are fully analysed by \cite{SC06.1} (hereafter [S06]),
who aim at compiling a list of candidate dark-matter halos which is as
complete as possible. They are selected with the aperture mass $M_\mathrm{ap}$
combined with a different statistic evaluating the probability for finding a
cluster with a given S/N at a particular position. The latter is also based on
$M_\mathrm{ap}$ ; see [S06] for details. They find 158 detections, $\sim50\%$
of which coincide with galaxy over-densities visible in the GaBoDS images.

In the present analysis based on the filter $\Psi$ optimised for
suppressing the noise, we reject all detections below a minimum
signal-to-noise ratio. This threshold is motivated by a statistical
analysis performed on the same data (see Sect.~\ref{sec:statistics}),
even if they coincide with known clusters or galaxy overdensities. In
this way, we obtain a list which is shorter than that obtained by
[S06], but contains only those detections which are statistically
reliable.

We also present a statistical comparison between the results obtained
with our optimal filter and with the two versions of the aperture
mass.

\subsection{Formalism for discrete data}

Since our estimation is performed on discrete sources, the integral in
Eqs.~(\ref{eq:mapt}) and (\ref{eq:estimate}) must be approximated by a
sum over the background galaxies,
\begin{equation} 
  A_\mathrm{est}(\vec\theta)=\frac{1}{n_\mathrm{g}}\, 
  \sum_i\epsilon_{\mathrm{t}i}(\vec\theta_i)\, 
  \Psi(|\vec\theta_i-\vec\theta|)\;.
\label{eq:discrsig}  
\end{equation}
The noise estimate in $A_\mathrm{est}$ is obtained from the variance
definition
$\sigma^2\equiv\left\langle(A_\mathrm{est}-A)^2\right\rangle$ as
\begin{equation}\label{eq:sigma}
  \sigma^2_{A_\mathrm{est}}(\vec\theta)= 
  \frac{1}{2n_\mathrm{g}^2}\,
  \sum_i|\epsilon_{\mathrm{t}i}(\vec\theta_i)|^2  
  \Psi^2(|\vec\theta_i-\vec\theta|)\;, 
\label{eq:discrnoise} 
\end{equation} 
where $n_\mathrm{g}$ is the number density of sources,
$\epsilon_{\mathrm{t}i}(\vec\theta_i)$ denotes the tangential
component of the $i$th galaxy's ellipticity relative to the line
connecting $\vec\theta_i$ and $\vec\theta$. Thus, the filtering is
performed in the real domain, which requires to Fourier back-transform
the filter function from
Eq.~(\ref{eq:optfilter}). Equation~(\ref{eq:sigma}) is a good
approximation where the local shear is negligible compared to the
intrinsic galaxy ellipticities, but it will typically fail near
clusters. This limitation can be overcome by substituting the variance
of the intrinsic galaxy ellipticities measured in the entire field.

Since the aperture mass and the optimal filter use the same linear
estimator, this discretisation is also used for the aperture mass, by
substituting $Q$ for $\Psi$ in Eqs.~(\ref{eq:discrsig}) and
(\ref{eq:discrnoise}).

\begin{figure*}[t]
   \centering
   \includegraphics[width=5.5cm]
 	{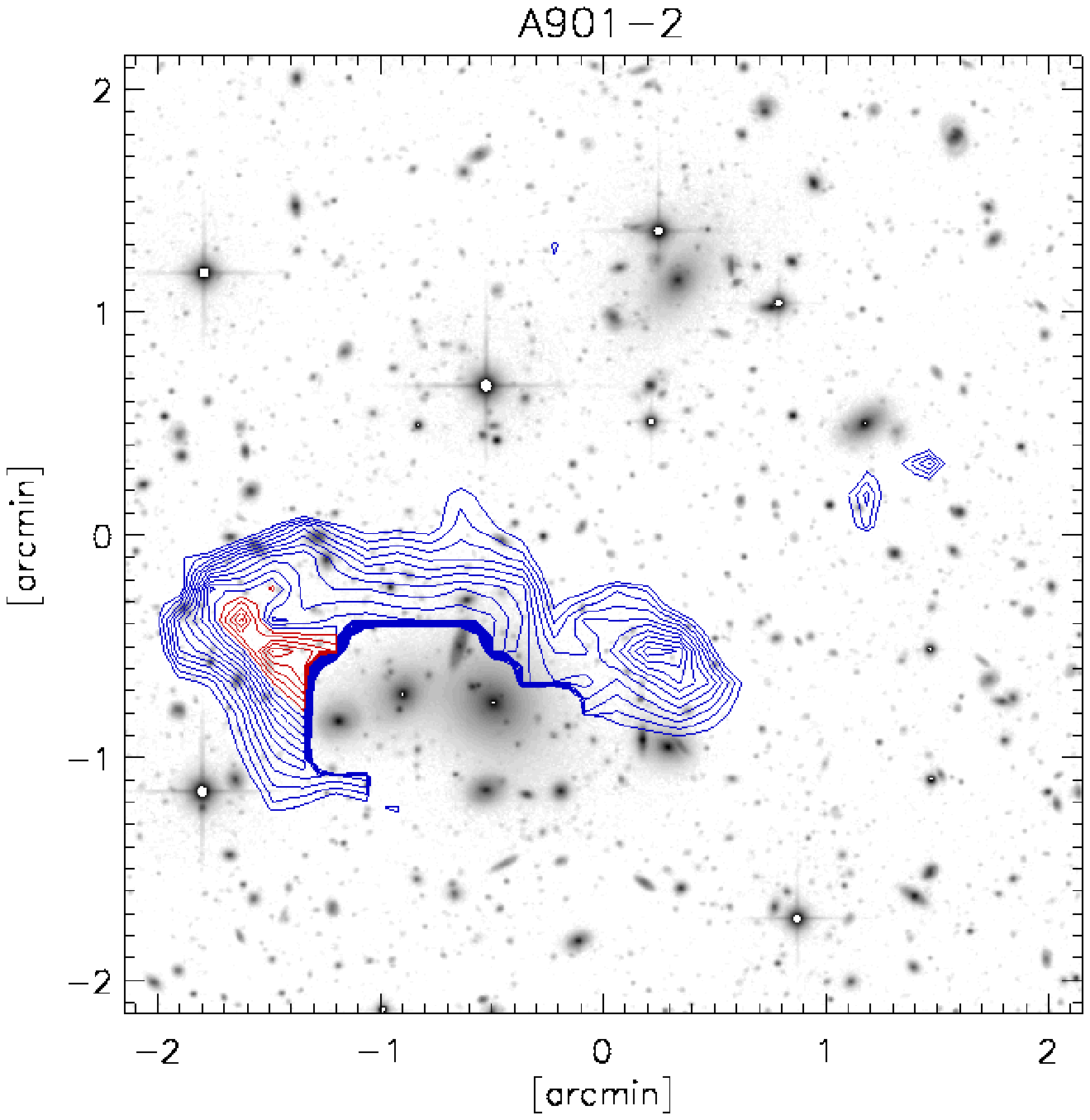}
   \hspace{-0.7cm}
   \includegraphics[width=5.5cm]
 	{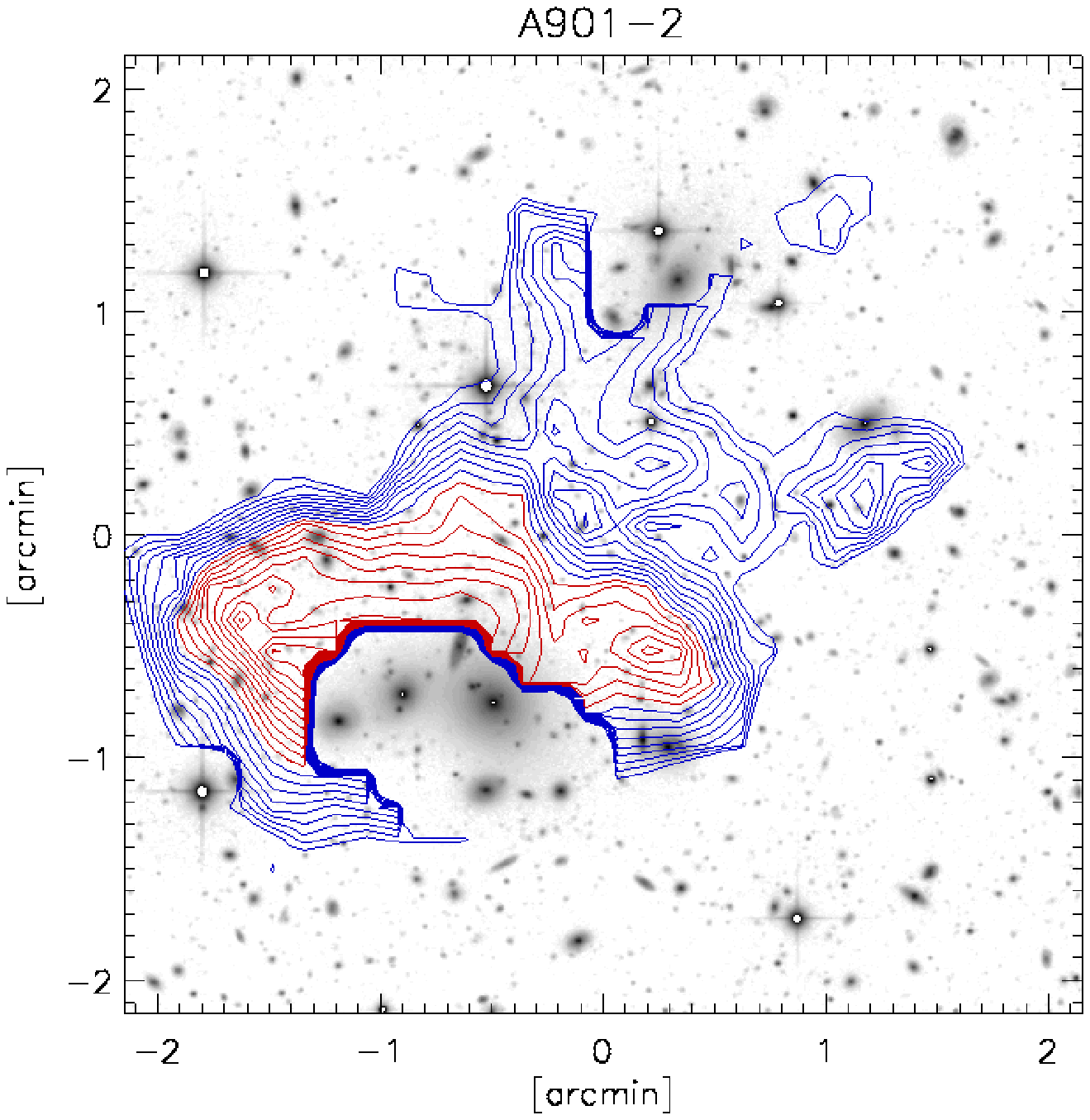}
   \hspace{-0.7cm}
   \includegraphics[width=5.5cm]
 	{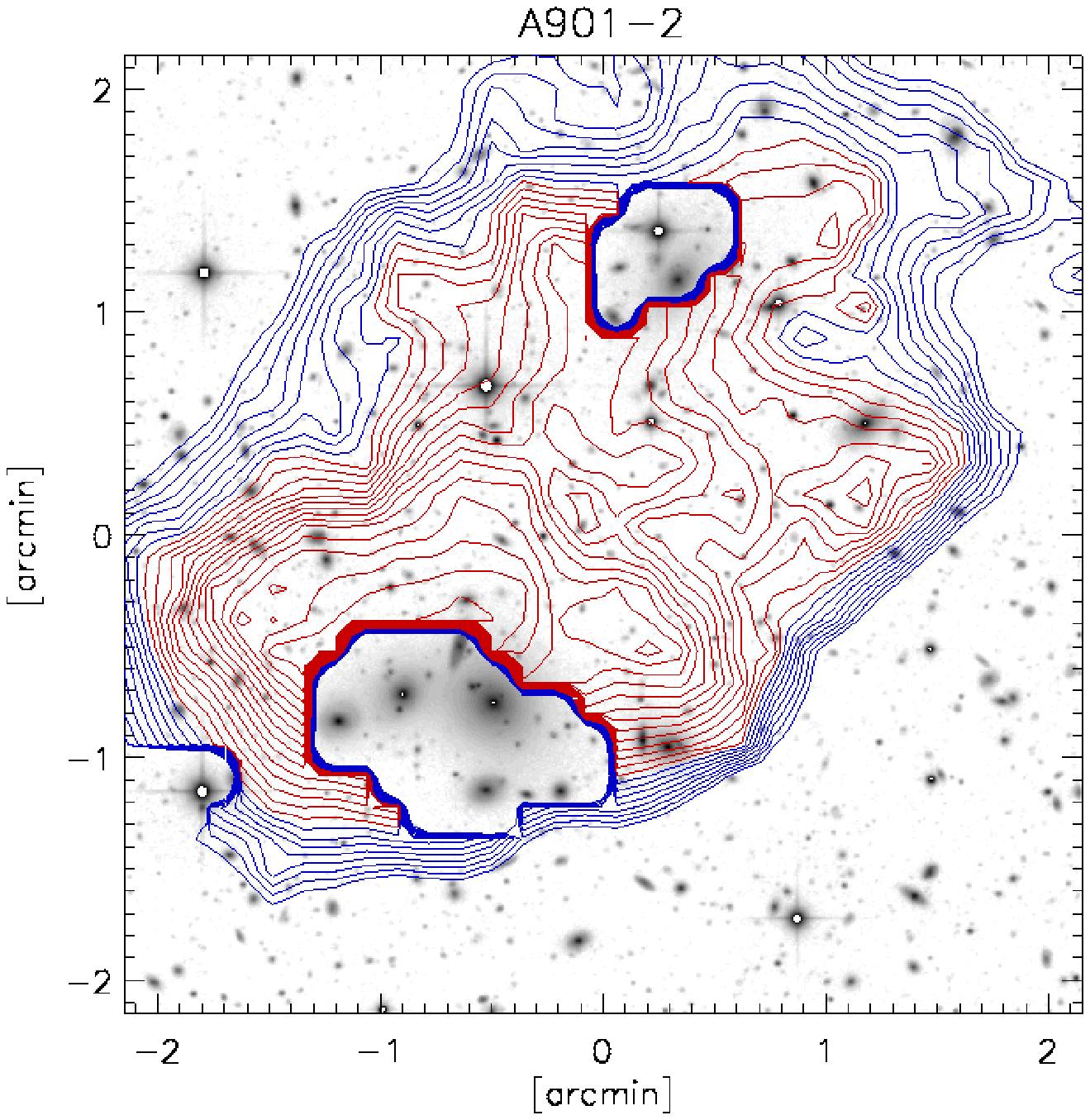}
\caption{ Example of a weak-lensing detection obtained with the
  optimal filter $A_\mathrm{opt}$ for the three different filter
  scales adopted in this study; $1'$, $2'$ and $4'$. The
  signal-to-noise iso-contours of the weak-lensing signal are
  superimposed here on the R-band image. The panel has a side length
  of $4.3'$, and the iso-contour lines start from $S/N=3$ with a step
  width of $0.1$. The red iso-contours are for $S/N>4.37$, $S/N>4.12$
  and $S/N>3.90$, respectively, and mark the reliable detections for
  each filter scale. Note the mask applied to remove bright galaxies
  and stars from the weak-lensing data.}
  \label{fig:images+SN}
\end{figure*}

\subsection{Signal-to-noise maps\label{sec:StoNmaps}}

Using Eqs.~(\ref{eq:discrsig}) and (\ref{eq:discrnoise}), we produce maps of
the signal-to-noise ratio for all our fields, henceforth called \emph{S-maps},
using the aperture masses $M_\mathrm{ap}$ and $M_\mathrm{oap}$ given by
Eqs.~(\ref{eq:ap_pol}) and (\ref{eq:ap_tanh}), and the optimal filter
$A_\mathrm{opt}$ given by Eq.~(\ref{eq:opt_filter}). The optimal filter $\Psi$
used in $A_\mathrm{opt}$ is computed for each field, because it has to be
optimised according to the observed galaxy number density. We use different
filter sizes to test the stability of the results achieved.

In order to investigate the noise properties of our data and to test the
reliability of the detections, we repeated the analysis on data catalogues in
which all individual shear measurements were rotated with a phase angle of
$45$ degrees, deleting any lensing signal contained in the data. The analysis
of these ``rotated'' data sets yields signal-to-noise maps similar to the
previous S-maps, but containing only noise due to the residual $B$-modes and
to the shot noise present in the data, and where no detections caused by
lensing can be found. We call these maps \emph{N-maps}.

We define the detections from the S- and N-maps as peaks above a given S/N
threshold and call them \emph{S-} and \emph{N-detections}, respectively. The
detections are automatically identified using a friends-of-friends (FF)
algorithm which finds groups of pixels on the S- and the N-maps connected with
a linking length of $1'$. This procedure avoids that S/N peaks associated with
the same distortion pattern are identified as different detections. In fact,
the signal given by a single halo can be fragmented by the noise, especially
for low signal-to-noise ratios and small filter scales, for which the shot
noise is more relevant.

We reject all fields where the noise given by the residual $B$-mode, as
detected in the N-maps, gives spurious detections which are not compatible
with the typical properties of the noise of our full data sample, i.e. with
signal-to-noise ratios and angular extentions belonging to extended tails in
the respective distributions. This selection has only a small impact on the
final detection sample. The selection of these fields depends on the filter
used, because of their different sensitivity on different scales. Our analysis
revealed that this contamination is larger for filters with a larger aperture,
as can be seen in Tab.~(\ref{tab:rejected_F}), where we list the rejected
fields.

In Sect.~(\ref{sec:statistics}) we discuss the statistical properties
of the detections, defining a threshold criterion which justifies the
reliability of the peaks described in Sect.~(\ref{sec:detections}).

\subsection{Statistics of the detections\label{sec:statistics}}

The statistical properties of the S- and N-detections, described in
Sect.~(\ref{sec:StoNmaps}), are shown in the upper and lower panels of
Fig.~(\ref{fig:cum_SN}), respectively. There, we plot the cumulative
distributions of the number of detections as a function of their maximum
signal-to-noise ratio. The curves are shown for $M_\mathrm{ap}$ (left panels),
$M_\mathrm{oap}$ (middle panels) and $A_\mathrm{opt}$ (right panels). Each
filter was used with three different scales; $1'$, $2'$ and $4'$ for
$M_\mathrm{ap}$ and $A_\mathrm{opt}$, and $5'$, $10'$ and $20'$ for
$M_\mathrm{oap}$.

All filters find many spurious N-detections, but the statistical
properties of data allow us to define a criterion to identify only the
most reliable detections. In fact, for $M_\mathrm{oap}$ and
$A_\mathrm{opt}$ only, the N-distribution drops faster than the
S-distribution, and for signal-to-noise ratios larger than a threshold
depending on the filter, no spurious detections are found. This
defines the signal-to-noise threshold to be used in selecting the
detections.

To better compare the S- and N-distributions for each filter, we plot in
Fig.~(\ref{fig:cum_SN-N}) their difference (upper panels) and the ratio
between the S- and the sum of the S- and the N-distributions (bottom
panels). The bottom panels show only the range between $0.5$ and $1$,
i.e.~where the number of S-detections is at least as high as the number of
N-detections. The figure clearly shows that the number of N-detections at high
signal-to-noise ratios obtained with $M_\mathrm{ap}$ (left panels) is
comparable or even larger than the number of S-detections except for an
aperture of $4'$. Consequently, this filter does not allow any distinction
between real detections and noise peaks, showing a strong dependence on the
chosen aperture. In comparison, the detections with $M_\mathrm{oap}$ and
$A_\mathrm{opt}$, shown in the centre and right panels respectively, have
substantially improved properties. Their S-detection distributions have a tail
extending towards high signal-to-noise ratios, rising above the number of
noise detections. As mentioned before, this shows that it is possible with
these filters to define a signal-to-noise threshold where the number of
N-detections drops below the number of S-detections. Detections above this
threshold define reliable peaks for all scales.

Applied to the GaBoDS data analysed here, even though $M_\mathrm{oap}$
performed less well than $A_\mathrm{opt}$, it returned almost
comparable results. This is not surprising because the filter shapes
are based on the same model for the halo density profiles. Thus, the
only decisive difference is that $A_\mathrm{opt}$ additionally
suppresses the noise contribution of the LSS.  The low number density
of galaxies in the GaBoDS survey and their low average redshift tend
to reduce this difference. However, this additional feature should be
of primary importance in deeper surveys such as the Cosmic Evolution
Survey\footnote{COSMOS: http://www.astro.caltech.edu/~cosmos}, whose
average source redshift is higher, increasing the amount of matter
projected along the line-of-sight and consequently the contamination
it causes.

\subsection{The significant weak-lensing
  detections\label{sec:detections}}

Since $A_\mathrm{opt}$ and $M_\mathrm{oapt}$ return similar results,
but the former has the further advantage of yielding the most reliable
and stable detections, we analyse only its results in detail.

Given the noise properties of our data sample and if we consider $\sim2$
spurious detections on the whole sample of about 18 square degrees an
acceptable contamination, the lower signal-to-noise threshold described in
Sect.~(\ref{sec:statistics}) is fixed at $4.37$, $4.12$ and $3.90$ for the
three adopted filter scales of $1'$, $2'$ and $4'$ respectively.

\begin{table*}[!t]
\caption{List of all reliable detections obtained with the optimal filter with
  scales of $1'$, $2'$ and $4'$. The adopted signal-to-noise thresholds are
  $4.37$, $4.12$ and $3.90$ according the filter scale. The columns give a
  preliminary name for the detections, their coordinates (2000.0), the
  signal-to-noise ratios obtained with the three filter scales, the number of
  galaxies per square arc minute in the respective field, the class of the
  detection (see the text for further explanation,
  Sect.~\ref{sec:detections}), the name of a corresponding object if present,
  and its redshift. The objects labeled [S06] are the
  detections obtained with the aperture mass by \cite{SC06.1}. The signal-to-noise
  ratios in brackets are below the signal-to-noise
  threshold.\label{tab:detect}}
\begin{tabular}{|l|ll|ccc|c|cll|}
  \hline
  ID & $\alpha$ (2000.0) & $\delta$ (2000.0) &
  opt $1'$ &opt $2'$ &opt $4'$ &n &class & object &z\\
  &  &  & $4.37<S/N$ &$4.12<S/N$ &$3.90<S/N$ & & & &\\
  \hline
  A901-1       &09 56 30 &-09 57 10 & 4.60         &4.91         &5.22         &25 &1&A901a       &0.16\\
  A901-2       &09 55 38 &-10 09 46 & 4.43         &4.21         &{\it (3.61)} &25 &3&J095538.2-101019  &0.169\\
  AM1-1        &03 53 26 &-49 42 49 & 4.86         &4.63         &4.05         &11 &6&023 [S06]   &--\\
  B8p2-1       &22 40 33 &-08 23 23 & {\it (3.99)} &4.19         &4.36         &11 &5&089 [S06]   &-- \\
  C04p2-1      &14 18 58 &-11 25 19 & {\it (4.34)} &4.13         &{\it (3.47)} & 8 &4&098 [S06]   &--\\
  CL1037-1243-1&10 38 24 &-12 32 38 & {\it (3.95)} &{\it (3.97)} &3.97         & 8 &4&--        &--\\
  CL1119-1129-1&11 18 22 &-11 27 12 & {\it (4.26)} &4.33         &4.01         & 9 &4&058 [S06] &-- \\
  Deep1a-1     &22 54 02 &-40 07 13 & 4.40         &4.34         &4.12         &14 &4&ESP 38      &0.151\\
  Deep1b-1     &22 50 25 &-40 05 55 & {\it (3.85)} &{\it (4.07)} &4.15         & 8 &5&--&--\\
  Deep3d-1     &11 18 45 &-21 35 57 & 4.68         &4.55         &4.27         &15 &6&060 [S06]          & --\\
  Deep3d-2     &11 17 37 &-21 34 16 & 4.55         &4.34         &3.96         &15 &4&057 [S06] & --\\
  F17\_P1-1    &14 26 38 &-34 46 08 & {\it (4.13)} &4.25         &4.12         & 7 &5&--                & --\\
  S11-1        &11 43 34 &-01 45 42 & {\it (3.91)} &4.15         &4.32         &18 &1&A1364             &0.107 \\
  SGP-1        &00 45 21 &-29 23 31 & {\it (3.89)} &{\it (3.92)} &3.97         &25 &1&LIL004521-292335  &0.257\\
  \hline
\end{tabular}
\end{table*}

According to this selection criterion, we identify the candidate
dark-matter halos from the data sample, keeping the contamination by
spurious detections under control by definition of the filter. Doing
so, we obtain a reliable sample consisting of $6$, $11$ and $12$
detections for the three apertures, respectively. These detections are
not independent, i.e.~most of them are achieved with more than one
aperture size, confirming the stability of our filter. Thus, our
selected sample of $14$ detections is composed as follows:

\begin{itemize}
\item 5 previously known clusters;
\item 4 of class 4;
\item 5 of class 5-6.
\end{itemize}

Five of our detections correspond to known clusters listed in the
SIMBAD Astronomical Database: Abell~1364, Abell~901,
J095538.2$-$101019, ESP~38 and LIL004521$-$292335. The classification
of the other detections is given according to the galaxy concentration
observed in the images according to the following empirical classes:
(1) very clear over-density with more than 50 bright galaxies;
(2) clear over-density of 35-50 bright galaxies;
(3) still highly significant over-density of 25-35 galaxies;
(4) still significant over-density of 15-25 galaxies;
(5) very loose distribution of 5-15 galaxies;
(6) absence of any kind of galaxy over-density.
We use these categories to facilitate the comparison of our detections
with the GaBoDS images. We are aware that this scheme may be
questionable for the classification of galaxy clusters.

Table~(\ref{tab:detect}) lists all candidates with their characteristics. The
columns, from left to right, give a preliminary name for the detections, their
coordinates (2000.0), the signal-to-noise ratios achieved with the three
adopted filter scales, the number of galaxies per square arc minute of the
respective fields, the classification of the detection, the name of the
corresponding object (if known) and its redshift. The objects labeled
[S06] are the detections obtained by \cite{SC06.1} by using different
combinations of the aperture-mass. The signal-to-noise ratios in brackets are
lower than the thresholds of statistically reliable detections.

Recall that we rejected all detections below the thresholds even if
associated with known clusters. This is so for LIL004608.1$-$292340
and LIL004544.0$-$294757, which are detected at $3.76\sigma$ and
$3.83\sigma$, respectively. This is because we want to carry out a
blind search for candidate clusters based exclusively on weak lensing
data.

An example for a detection is shown in Fig.~(\ref{fig:images+SN})
where we plot the optical R-band image of the cluster Abell~901 with
iso-contours of the signal-to-noise ratio of the weak-lensing signal
superimposed. The three panels, with a side length of $4.3'$ each, are
obtained with the three filter scales adopted in this work. The
iso-contour lines start from $S/N=3$ with a step width of $0.1$. The
levels of the red iso-contours are at $S/N>4.37$, $S/N>4.12$ and
$S/N>3.90$, respectively, and mark the reliable detections for each
filter scale. To be conservative, we avoid to compute the
signal-to-noise ratio where the bright galaxies and stars were masked
from the data. The cluster Abell~901 is a massive structure at
redshift $z=0.16$. Due to its size, a filter optimised for larger
scales yields higher signal-to-noise ratios, and the extent of the
dark matter halo is better visible (see the right panel).

\section{Conclusions\label{sec:conclude_gabods}}

The application of our filter to the GaBoDS weak-lensing survey
confirms the expectations on its performance raised by \cite{MAT04.2}
based on numerical simulations. Our filter is more stable than the
aperture mass against changes of the filter size, thus considerably
simplifying the interpretation of data. It has better statistical
properties compared to the aperture mass given by
Eq.~(\ref{eq:ap_pol}), yielding more reliable results. Although our
filter and the optimised aperture mass given by Eq.~(\ref{eq:ap_tanh})
perform comparably on the GaBoDS data, the LSS suppression
characterising our filter will be primarily important for deeper
observations, for which the average background-galaxy redshift will be
larger than in GaBoDS and the LSS will not be negligible.

We measured in our data sample a large contamination from residual
$B$-modes, but a statistical analysis of the noise properties allowed
us to define a criterion to select a sample of reliable detections. We
emphasise that a data reduction procedure which minimises the residual
$B$-mode is of fundamental importance for the detection of new
cluster-sized dark-matter halos.

We rejected all detections below the thresholds even if associated with know
clusters, because we wish to search blindly for candidate clusters based
exclusively on weak-lensing data. On the $19.6$ square degrees covered by the
GaBoDS data, we found $14$ detections with a sufficiently high signal-to-noise
ratio, $5$ of which are known clusters, 4 are associated with concentrations
of galaxies visible in our data, and 5 detections are not associated with any
visible concentration of galaxies. Deep optical and $X$-ray follow-ups of the
$9$ unknown detections should be performed to clarify their nature.

\acknowledgements{This work has been partially supported by the Vigoni
  program of the German Academic Exchange Service (DAAD) and Conference of
  Italian University Rectors (CRUI). We are grateful to Peter Schneider for
  helpful discussions.}

\end{document}